\documentclass[english,english,pra,english,preprint,amsmath,amssymb,aps,longbibliography,showkeys,titlepage]{revtex4-2}
\usepackage{lmodern}
\usepackage{lmodern}
\usepackage[T1]{fontenc}
\usepackage[utf8]{luainputenc}
\usepackage{array}
\usepackage{amsmath}
\usepackage{amsthm}
\usepackage{amssymb}
\usepackage{graphicx}
\usepackage{esint}

\makeatletter

\providecommand{\tabularnewline}{\\}

\usepackage{amsthm}\usepackage{latexsym}\usepackage{bm}\usepackage{amsfonts}

\usepackage{babel}
\usepackage{hyperref}
\usepackage{enumitem}

\usepackage{hyperref}
\hypersetup{
	breaklinks = true,
    colorlinks = true,
    citecolor = {blue},
	urlcolor = {blue},
	linkcolor = {blue}
}

\makeatother

\usepackage{babel}
\begin{document}
\title{Algebraic discrete quantum harmonic oscillator with dynamic resolution
scaling }
\author{Michael Q. May}
\email{mqmay@princeton.edu}

\affiliation{Princeton Plasma Physics Laboratory and Department of Astrophysical
Sciences, Princeton University, Princeton, NJ 08540, U.S.A}
\author{Hong Qin}
\email{hongqin@princeton.edu}

\affiliation{Princeton Plasma Physics Laboratory and Department of Astrophysical
Sciences, Princeton University, Princeton, NJ 08540, U.S.A}
\begin{abstract}
We develop an algebraic formulation for the discrete quantum harmonic
oscillator (DQHO) with a finite, equally-spaced energy spectrum and
energy eigenfunctions defined on a discrete domain, which is known
as the $\mathfrak{su}(2)$ or Kravchuk oscillator. Unlike previous
approaches, ours does not depend on the discretization of the Schrödinger
equation and recurrence relations of special functions. This algebraic
formulation is endowed with a natural $\mathfrak{su}(2)$ algebra,
each finite dimensional irreducible representation of which defines
a distinct DQHO labeled by its resolution. In addition to energy ladder
operators, the formulation allows for resolution ladder operators
connecting all DQHOs with different resolutions. The resolution ladder
operators thus enable the dynamic scaling of the resolution of finite
degree-of-freedom quantum simulations. Using the algebraic DQHO formalism,
we are able to rigorously derive the energy eigenstate wave functions
of the QHO in a purely algebraic manner without using differential
equations or differential operators, which is impossible in the continuous
or infinite discrete setting. The coherent state of the DQHO is constructed,
and its expected position is proven to oscillate as a classical harmonic
oscillator. The DQHO coherent state recovers that of the quantum harmonic
oscillator at large resolution. The algebraic formulation also predicts
the existence of an inverse DQHO that has no known continuous counterpart. 
\end{abstract}
\maketitle

\section{Introduction}

Without regard to complications including q-deformation, \emph{discrete
quantum harmonic oscillator} refers to the class of\emph{ }Meixner
oscillators\emph{ }including the Meixner, Kravchuk, and Charlier oscillators.
The wave functions of each of these oscillators become Hermite functions
(the energy eigenstates of the quantum harmonic oscillator) in certain
limits, and their energy spectra are equally spaced. The limiting
relationships between these oscillators has been documented \citep{atakishiev98}.
In this work, we will consider the Kravchuk oscillator, which is the
only finite dimensional discrete quantum harmonic oscillator (DQHO)
with equally-spaced energy eigenvalues. It has been independently
derived at least twice \citep{atakishiev90,lorente97} by discretizing
the Schrödinger equation for the quantum harmonic oscillator (QHO).
However, perhaps owing to the extensive recurrence relations and identities
of special functions present in both derivations and most discussions
of the oscillator, it remains relatively obscure. For example, the
Kravchuk oscillator has also been unknowingly rederived in the context
of transferring information along quantum spin chains \citep{christandl,shi_sun}.
Also, despite the growing interest in quantum computing, where discreteness
and finitude is an advantage, the Kravchuk oscillator has never been
simulated on a quantum computer (though the closely related three-wave
interaction has \citep{shiPhysRevA}). Indeed, quantum computation
of the QHO has been restricted to non-structure preserving, finite-difference
methods \citep{miceli19,jain21}.

Because the Kravchuk functions are the finite, discrete variable versions
of the Hermite functions, previous derivations of the Kravchuk oscillator
have taken the Kravchuk wavefunctions as an \emph{ansatz} \citep{lorente01,lorente97,atakishiev90,atakishiev98,atakishiyev08}.
Finding the Hamiltonian, position and momentum eigenfunctions, and
raising and lowering operators is then an exercise in cleverly applying
known recurrence relations. Though straightforward, this approach
based on special functions can obscure important structures which
arise in Kravchuck oscillator. For example, the spectrum was found
to have an $\mathfrak{su}(2)$ structure, and for this reason, the
Kravchuck oscillator is also known as a $\mathfrak{su}(2)$ oscillator.
However, rather than proving fundamental to the derivation, the $\mathfrak{su}(2)$
algebraic structure underlying the oscillator was discovered only
after a complex process of defining raising and lowering operators
using hypergeometric functions. And, rather than being a dynamic quantity,
the resolution of the discrete oscillator is simply taken as a fixed
quantity at the start of the derivation. 

In the present study, we develop an algebraic formulation for the
DQHO that has a natural $\mathfrak{su}(2)$ structure and is fully
equivalent to the Kravchuk oscillator at a fixed dimension. The dimension
of a DQHO, also referred to as the resolution, is the number of the
discrete energy eigenstates or grid points. In addition to recreating
previous results in a simpler way, the algebraic formulation furnishes
new structures that can dynamically change the resolution of the DQHO.
In particular, in addition to energy raising and lowering operators,
we will build resolution raising and lowering operators which connect
DQHOs with different resolutions. The formulation is developed from
the minimally coupled Hamiltonian for two quantum harmonic oscillators,
defined in Eq.\,(\ref{eq:dqho_ham}). 

In Section \ref{sec:Algebraic-Derivation-of}, we show that the Hamiltonian
has an equally-spaced spectrum and construct quadratic energy raising
and lowering operators, which, with $H$, form the complexified $\mathfrak{su}(2)$
Lie algebra. It is then proven that the spectrum of the Hamiltonian
consists of all finite dimensional irreducible representations of
the $\mathfrak{su}(2)$ algebra. Dimension raising and lowering operators
naturally arise and allow one to move between different representations.
We show that each finite dimensional irreducible representation of
the $\mathfrak{su}(2)$ algebra in the full spectrum defines a DQHO
that recovers the Kravchuk oscillator at the same dimension. We also
find that there exist inherent nonrenormalizable energy eigenstates
in the $\mathfrak{su}(2)$ oscillator and give a physical interpretation
facilitated by the algebraic formulation. In Section \ref{sec:Coherent-States},
we discuss the definition of coherence in a finite, discrete context
and find coherent states by application of a discrete analog of the
displacement operator to the ground state of the algebraic DQHO. Finally,
in Section \ref{sec:Discussion} we summarize and discuss extensions. 

\section{Algebraic formulation of the DQHO\label{sec:Algebraic-Derivation-of}}

We begin with the minimally coupled Hamiltonian for two quantum harmonic
oscillators, 

\begin{equation}
H=\frac{1}{2}\left(A_{1}^{\dagger}A_{1}+A_{2}^{\dagger}A_{2}\right)-\frac{1}{2}\left(A_{1}A_{2}^{\dagger}+A_{1}^{\dagger}A_{2}\right),\label{eq:dqho_ham}
\end{equation}
where $A_{1}^{\dagger}$, $A_{1}$, $A_{2}^{\dagger}$, and $A_{2}$
are creation and annihilation operators satisfying the usual commutation
relations, 

\begin{equation}
[A_{i},A_{j}^{\dagger}]=\delta_{ij}.\label{eq:AiAj}
\end{equation}
The Hamiltonian is composed of two dynamic parts,

\begin{equation}
H=\frac{1}{2}S+H_{I},
\end{equation}
with the total number of quanta shared by the oscillators
\begin{align}
S & \equiv N_{1}+N_{2}\,,\label{eq:S}\\
N_{1} & \equiv A_{1}^{\dagger}A_{1}\,,N_{2}\equiv A_{2}^{\dagger}A_{2}\,,
\end{align}
and the interaction term defined by 
\begin{equation}
H_{I}=-\frac{1}{2}\left(A_{1}A_{2}^{\dagger}+A_{1}^{\dagger}A_{2}\right).\label{eq:Hi}
\end{equation}
Both the number of quanta $S$ and the interaction term $H_{I}$ are
Hermitian, and the set $\{H,S,H_{I}\}$ is mutually commuting. Because
the eigenvalue of $S$ will end up being one less than than the number
of the lattice points, or resolution, of the discrete system, $S$
will be called the resolution operator. Denote the eigenstates of
energy $H$ and resolution $S$ by $|n,s\rangle,$ where $n$ and
$s$ are the eigenvalues of $H$ and $S$, respectively. We will refer
to $|n,s\rangle$ as the energy-resolution eigenstates. 

The interaction term, $H_{I}$, appears in several different physical
systems. It is used as the interacting Hamiltonian between two sites
in the Bose-Hubbard mode. It can also be viewed as a special case
of the quantum three wave system when one wave is strong and stationary
\citep{Shi2017,Shi2018,Shi2021,shiPhysRevA,may_qin,may_qin_nonlinear_three_wave}.
In certain contexts of quantum optics, it has been referred to as
a beam splitter Hamiltonian. Another well-known setting where it appears
is in Schwinger's construction of the angular momentum algebra from
two uncoupled QHOs governed by the Hamiltonian $S=A_{1}^{\dagger}A_{1}+A_{2}^{\dagger}A_{2}$
\citep{schwinger}. However, in Schwinger's construction, $H_{I}$
appears as an angular momentum, instead of as part of the Hamiltonian
of the system. 

Our first goal in this section is to show that the spectrum of the
Hamiltonian $H$ consists of all finite dimensional irreducible representations
of the $\mathfrak{su}(2)$ Lie algebra, each of which defines a DQHO
with a resolution equal to its dimension. 

The equations of motion according to Eq. (\ref{eq:dqho_ham}) are 

\begin{equation}
[H,A_{1}]=\frac{1}{2}\left(A_{2}-A_{1}\right),\thinspace\thinspace[H,A_{2}]=\frac{1}{2}\left(A_{1}+A_{2}\right).\label{eq:EoM_A1}
\end{equation}
For further analysis, we can trade $A_{1}$ and $A_{2}$ for operators
$B_{1}$ and $B_{2}$ defined as follows,
\begin{equation}
B_{1}\equiv A_{1}+A_{2},\thinspace\thinspace B_{2}\equiv A_{1}-A_{2}.\label{eq:bc}
\end{equation}
Note that as with $A_{1}$ and $A_{2}^{\dagger}$, $B_{1}$ commutes
with $B_{2}^{\dagger}$. This allows us to write 
\begin{equation}
H=\frac{1}{2}B_{2}^{\dagger}B_{2},
\end{equation}
obviating the commutators
\begin{equation}
[H,B_{1}]=0,\thinspace\thinspace[H,B_{2}]=-B_{2}.
\end{equation}
Next we define the following quadratic operators in terms of $B_{1}$
and $B_{2}$,

\begin{equation}
D\equiv B_{1}^{\dagger}B_{2},\thinspace\thinspace D^{\dagger}\equiv B_{1}B_{2}^{\dagger}.\label{eq:DF}
\end{equation}
Calculation shows that $D$ and $D^{\dagger}$, while commuting with
$S$, act as lowering and raising operators of the energy, 
\begin{equation}
[H,D]=-D,\thinspace[H,D^{\dagger}]=D^{\dagger}.
\end{equation}
Similarly, the operators $B_{1}$ and $B_{1}^{\dagger}$, which commute
with $H$, are the raising and lowering operators of the resolution,
\begin{equation}
[S,B_{1}]=-B_{1},\thinspace\thinspace[S,B_{1}^{\dagger}]=B_{1}^{\dagger}.
\end{equation}
Commutation relations for the algebra of \{$H$, $S$, $B_{1}$, $B_{2}$,
$B_{1}^{\dagger}$, $B_{2}^{\dagger}$, $I$\} are summarized in Table
\ref{table_1}. 
\begin{table}
\begin{tabular}{>{\centering}p{1.6cm}|>{\centering}p{1.6cm}>{\centering}p{1.6cm}>{\centering}p{1.6cm}>{\centering}p{1.6cm}>{\centering}p{1.6cm}>{\centering}p{1.6cm}>{\centering}p{1.6cm}}
$[\downarrow,\rightarrow]$ & $H$ & $S$ & $B_{1}$ & $B_{1}^{\dagger}$ & $B_{2}$ & $B_{2}^{\dagger}$ & $I$\tabularnewline
\hline 
$H$ & $0$ & $0$ & $0$ & $0$ & $-B_{2}$ & $B_{2}^{\dagger}$ & $0$\tabularnewline
$S$ & $0$ & $0$ & $-B_{1}$ & $B_{1}^{\dagger}$ & $-B_{2}$ & $B_{2}^{\dagger}$ & $0$\tabularnewline
$B_{1}$ & $0$ & $B_{1}$ & $0$ & $2I$ & $0$ & $0$ & $0$\tabularnewline
$B_{1}^{\dagger}$ & $0$ & $-B_{1}^{\dagger}$ & $-2I$ & $0$ & $0$ & $0$ & $0$\tabularnewline
$B_{2}$ & $B_{2}$ & $B_{2}$ & $0$ & $0$ & $0$ & $2I$ & $0$\tabularnewline
$B_{2}^{\dagger}$ & $-B_{2}^{\dagger}$ & $-B_{2}^{\dagger}$ & $0$ & $0$ & $-2I$ & $0$ & $0$\tabularnewline
$I$ & $0$ & $0$ & $0$ & $0$ & $0$ & $0$ & $0$\tabularnewline
\end{tabular}\caption{Commutation relations for $H$, $S$, $B_{1}$, $B_{2}$, $I$, and
their complex conjugates.}
\label{table_1}
\end{table}

We now need to look more closely at the interaction term $H_{I}$
to determine how the resolution operator $S$ gets its name. The interaction
term $H_{I}$ has equations of motion 

\begin{equation}
[H_{I},A_{1}]=\frac{1}{2}A_{2},\thinspace\thinspace[H_{I},A_{2}]=\frac{1}{2}A_{1}.\label{eq:A1H-2}
\end{equation}
It is interesting to note that by transforming $H_{I}\rightarrow2H_{I}$
and $A_{1}\rightarrow-iA_{1}$, these take the same form as those
for the QHO with the canonically commuting $p$ and $q$ replaced
by the mutually commuting $A_{1}$ and $A_{2}$. In the $B_{1}$,
$B_{2}$ basis, these commutators become
\begin{equation}
[H_{I},B_{1}]=\frac{1}{2}B_{1},\,\,\thinspace\thinspace[H_{I},B_{2}]=-\frac{1}{2}B_{2}.
\end{equation}
We next find that the energy raising and lowering operators are also
raising and lowering operators of $H_{I}$,
\begin{equation}
[H_{I},D]=-D,\thinspace[H_{I},D^{\dagger}]=D^{\dagger}.
\end{equation}
These results are summarized in Table \ref{d_tab}. 
\begin{table}
\begin{tabular}{>{\centering}p{1.6cm}|>{\centering}p{1.6cm}>{\centering}p{1.6cm}>{\centering}p{1.6cm}>{\centering}p{1.6cm}>{\centering}p{1.6cm}>{\centering}p{1.6cm}}
$[\downarrow,\rightarrow]$ & $H_{I}$ & $D$ & $D^{\dagger}$ & $B_{1}$ & $B_{1}^{\dagger}$ & $I$\tabularnewline
\hline 
$H_{I}$ & $0$ & $-D$ & $D^{\dagger}$ & $\frac{1}{2}B_{1}$ & $-\frac{1}{2}B_{1}^{\dagger}$ & $0$\tabularnewline
$D$ & $D$ & $0$ & $8H_{I}$ & $-2B_{2}$ & $0$ & $0$\tabularnewline
$D^{\dagger}$ & $-D^{\dagger}$ & $-8H_{I}$ & $0$ & $0$ & $2B_{2}^{\dagger}$ & $0$\tabularnewline
$B_{1}$ & $-\frac{1}{2}B_{1}$ & $2B_{2}$ & $0$ & $0$ & $2I$ & $0$\tabularnewline
$B_{1}^{\dagger}$ & $\frac{1}{2}B_{1}^{\dagger}$ & $0$ & $-2B_{2}^{\dagger}$ & $-2I$ & $0$ & $0$\tabularnewline
$I$ & $0$ & $0$ & $0$ & $0$ & $0$ & $0$\tabularnewline
\end{tabular}\caption{Commutation relations for $H_{I}$, $D$, $B_{1}$, $I$, and their
complex conjugates.}
\label{d_tab}
\end{table}

Table \ref{d_tab} also highlights an important structure---$H_{I},$
$D$, and $D^{\dagger}$ form the complexification of the Lie algebra
$\mathfrak{su}(2)$. Scaling $H_{I} \rightarrow2H_{I}$, $D\rightarrow\frac{1}{2}D$,
and $D^{\dagger}\rightarrow\frac{1}{2}D^{\dagger}$ makes this obvious.
The quadratic Casimir $\Omega$ of $\mathfrak{su}(2)$, which indexes
the irreducible representations of the algebra according to its eigenvalue
$\omega$, is $\Omega=DD^{\dagger}+4H_{I}^{2}+4H_{I}$. The eigenvalue
\textbf{$\omega$ }of $\Omega$ is related to the dimension $d$ of
the corresponding irreducible representation via $\omega=(d-1)(d+1)$.
As usual, the algebra $\mathfrak{su}(2)$ itself does not select a
particular representation from all of its possible representations.
But, for the interaction term $H_{I}$ defined Eq.\,(\ref{eq:Hi}),
\begin{eqnarray}
\Omega=DD^{\dagger}+4H_{I}^{2}+4H_{I} & = & S(S+2).\label{eq:Omega}
\end{eqnarray}
This implies that the eigenvalues $s$ of $S$ determine the dimension
of the representation by $d=s+1$, which will be shown later to be
the number of lattice points, i.e. the resolution, of the wave functions
of the algebraic DQHO. 

For a fixed resolution, the eigenvalues of $H$ and $H_{I}$ will
differ only by a constant value, so $S$ also determines the number
of energy eigenstates of $H$ for a particular resolution. Thus, the
full spectrum of $H$ consists of sets of $s$ equally spaced eigenvalues,
for each $s\in\mathbb{N}$, corresponding to the irreducible representations
of $\mathfrak{su}(2).$

\subsection{Ladder operators in the energy-resolution basis}

In the energy-resolution basis, the normalization factors of the energy
raising and lowering operators are given by, 

\begin{eqnarray}
D^{\dagger}|n,s\rangle & = & 2\sqrt{s-n}\sqrt{n+1}|n+1,s\rangle,\label{eq:eng_raise}\\
D|n,s\rangle & = & 2\sqrt{s-n+1}\sqrt{n}|n-1,s\rangle.\label{eq:eng_lower}
\end{eqnarray}
An explanation of the process for finding these and other normalizations
can be found in Appendix A. These normalizations imply there is a
barrier to raising the energy to or lowering the energy from $n=0$
and a similar barrier for $n=s$. Between the energy barriers, for
a fixed value of $s$, there are therefore $s+1$ energy eigenstates
accessible to the energy lowering and raising operators. This number
of eigenstates is the dimension of the irreducible representation
of the $\mathfrak{su}(2)$ algebra labeled by $s$. 

The normalization factors for the resolution raising and lowering
operators are given by,
\begin{eqnarray}
B_{1}^{\dagger}|n,s\rangle & = & \sqrt{2(s-n+1)}|n,s+1\rangle,\\
B_{1}|n,s\rangle & = & \sqrt{2(s-n)}|n,s-1\rangle.
\end{eqnarray}
Because the resolution operators do not affect the energy eigenstates,
with proper normalization their application leaves quantum superpositions
untouched, effectively allowing for consistent, dynamic resolution
scaling of quantum simulations. 

\begin{figure}
\includegraphics[scale=0.5]{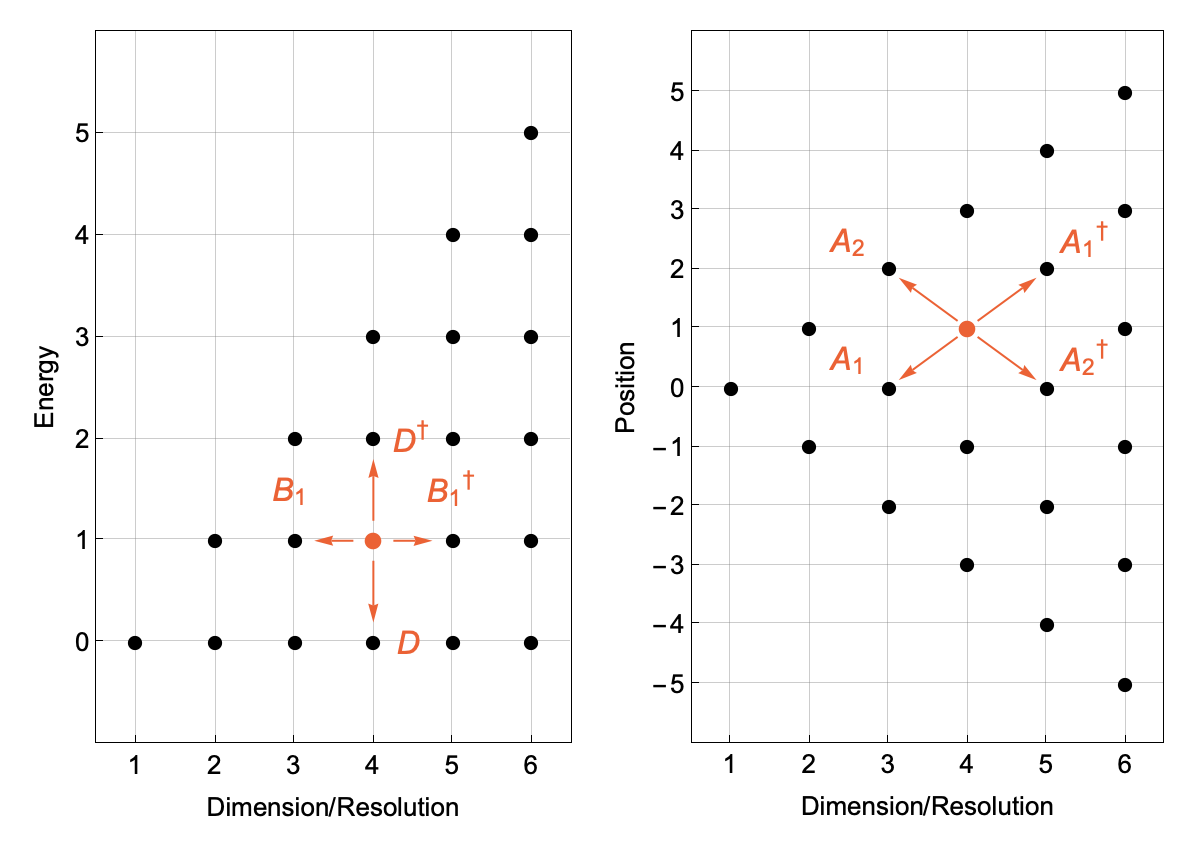}\caption{Actions of raising and lowering operators on energy-resolution and
position-resolution bases. Each dot represents an eigenstate, and
the arrows indicate how operators transform eigenstates. The arrows
are translationally invariant in each diagram. \emph{Left}: Action
of the $D$, $D^{\dagger}$, $B_{1}$, and \textbf{$B_{1}^{\dagger}$}
operators on the energy-resolution basis. \emph{Right}: Action of
the $A_{1}$, $A_{1}^{\dagger}$, $A_{2}$, $A_{2}^{\dagger}$ operators
on the position-resolution basis.}

\label{fig:operator_actions}
\end{figure}

How can the resolution operators not affect quantum superpositions
if they do not always commute with the energy ladder operators? Namely,
while $[D,B_{1}^{\dagger}]=[D^{\dagger},B_{1}]=0$, $[D,B_{1}]\not=0$
and $[D^{\dagger},B_{1}^{\dagger}]\not=0.$ This is a natural result
of the non-unitarity of the energy and resolution operators. For example,
since the energy raising operator destroys the highest energy state,
it will matter whether or not the resolution is increased before or
after the energy raising operator is applied. Similarly, since the
resolution lowering operator also destroys the highest energy state,
lowering its energy before lowering the resolution will have a different
result than attempting to lower the energy after. This can be seen
clearly in Fig. \ref{fig:operator_actions}, where the actions of
the $D$, $D^{\dagger}$, $B_{1}$, and $B_{1}^{\dagger}$ operators
in the energy-resolution basis are shown alongside the actions of
the $A_{1}$, $A_{2}$, $A_{1}^{\dagger}$, and $A_{2}^{\dagger}$
operators in the position-resolution basis, which is discussed below.

\subsection{Position and momentum operators}

By analogy with the QHO, we may define a position operator
\begin{equation}
X\equiv\frac{1}{2}\left(D^{\dagger}+D\right)=2N_{1}-S,\label{eq:X}
\end{equation}
and a momentum operator 
\begin{equation}
P\equiv\frac{i}{2}(D^{\dagger}-D),
\end{equation}
in terms of the energy raising and lowering operators. The equations
of motion of the QHO are satisfied by $X$ and $P$:
\begin{align}
[H,X] & =-iP,\\{}
[H,P] & =iX.
\end{align}
Note that the second equation of Eq.\,(\ref{eq:X}) implies that
the eigenvalues of position $x\in\{-s,-s+2,\dots,s\}$. However, as
with $D$ and $D^{\dagger}$, $H$ may not be written in terms of
$X$ and $P$ in the same way as it can for the QHO. Note that $X^{2}+P^{2}\propto D^{\dagger2}+D^{2}\not\propto H$. 

In the energy-resolution basis, we can calculate the expectation values
of $X$ and $P$. An arbitrary state of fixed dimension $\Psi_{s}$
in this basis is
\begin{equation}
\Psi_{s}=\sum_{j=0}^{s}d_{j}\left|j,s\right\rangle 
\end{equation}
with $d_{j}$ the complex amplitude of the energy eigenmode $|j\rangle$.
Then, 
\begin{equation}
\left\langle \Psi_{s}|X|\Psi_{s}\right\rangle =2\sum_{j=0}^{s-1}\text{Re}\left(d_{j}^{*}d_{j+1}\right)\sqrt{s-j}\sqrt{j+1}
\end{equation}
and 
\begin{equation}
\left\langle \Psi_{s}|P|\Psi_{s}\right\rangle =2\sum_{j=0}^{s-1}\text{Im}\left(d_{j}^{*}d_{j+1}\right)\sqrt{s-j}\sqrt{j+1}.
\end{equation}

\subsection{Position-resolution basis}

The Hilbert space spanned by the spectrum of $H$ has a different
basis indexed by quantum numbers $n_{1}$, $n_{2}$, and $s$, the
eigenvalues of the operators $N_{1}$, $N_{2}$, and $S$, respectively.
Because $s=n_{1}+n_{2}$, eigenstates in this basis can be labeled
by $n_{1}$ and $s$ only and will be denoted by $\psi_{n_{1},s}$.
This basis will be referred to as the position-resolution basis because
$\psi_{n_{1},s}$ is an eigenstate of the position operator $X$ defined
by Eq.\,(\ref{eq:X}). 

In the position-resolution basis, the expressions of $A_{1},A_{1}^{\dagger},A_{2},A_{2}^{\dagger}$
are
\begin{align}
A_{1}\psi_{n_{1},s}=\sqrt{n_{1}}\psi_{n_{1}-1,s-1},\thinspace\,\,\, & A_{1}^{\dagger}\psi_{n_{1},s}=\sqrt{n_{1}+1}\psi_{n_{1}+1,s+1}\thinspace,\\
A_{2}\psi_{n_{1},s}=\sqrt{s-n_{1}}\psi_{n_{1},s-1},\thinspace\,\,\, & A_{2}^{\dagger}\psi_{n_{1},s}=\sqrt{s+1-n_{1}}\psi_{n_{1},s+1}\thinspace,
\end{align}
from which the expressions of all other operators can be calculated.
For example, the expression of the Hamiltonian is

\begin{align}
H\psi_{n_{1},s} & =(\frac{1}{2}S+H_{I})\psi_{n_{1},s}\nonumber \\
 & =\frac{1}{2}\left(N_{1}+N_{2}\right)\psi_{n_{1},s}-\frac{1}{2}\left(A_{1}A_{2}^{\dagger}+A_{1}^{\dagger}A_{2}\right)\psi_{n_{1},s}\nonumber \\
 & =\frac{s}{2}\psi_{n_{1},s}+\sqrt{n_{1}}\sqrt{s-n_{1}+1}\psi_{n_{1}-1,s}+\sqrt{n_{1}+1}\sqrt{s-n_{1}}\psi_{n_{1}+1,s}.\label{eq:H}
\end{align}
Equation (\ref{eq:H}) shows again that in this basis, the eigenstates
of the Hamiltonian must necessarily have dimension $d=s+1$, since
the Hamiltonian spreads states from $\psi_{0,s}$ to $\psi_{s,s}$.
Specifically, in this basis with a fixed dimension, $H$ is a symmetric,
tridiagonal matrix where
\begin{gather}
H_{ij}=\delta_{i,j}s+\delta_{i+1,j}h_{i-1}+\delta_{i,j+1}h_{j-1},\label{eq:ham_mat}
\end{gather}
and
\begin{equation}
h_{i}=\sqrt{(s-i)(i+1)}.
\end{equation}
This matrix representation of the Hamiltonian is a known result for
the $\mathfrak{su}(2)$ structure of the Kravchuk DQHO \citep{atakishiev90,lorente97}.
It also appears in models of fermionic spin chains \citep{christandl,shi_sun},
the Hanh oscillator \citep{jafarov}, and the q-deformed harmonic
oscillator \citep{atakishiyev_01}. However, it should be emphasized
that the correspondence with these previous models is only for a fixed
number of lattice points, i.e. resolution. Previous models for the
DQHO do not provide an internally consistent method to specify and
change the resolution, which is one of the tools needed to calculate
the eigenstates in the position-resolution basis. This explains why
the eigenmode structures in previous DQHO models have to depend on
specific, non-trivial mathematical properties of special functions. 

\subsection{DQHO}

As shown above, for each value of $s$, the $s+1$ energy eigenstates
$|n,s\rangle$ furnish an $(s+1)$-dimensional irreducible representation
of the $\mathfrak{su}(2)$ algebra of $H_{I},$ $D^{\dagger}$, and
$D.$ In the present study, we will call the $s+1$ energy eigenstates
$|n,s\rangle$ an $(s+1)$-dimensional DQHO because they also furnish
an $(s+1)$-dimensional approximation of the QHO. The validity of
this approximation can be justified from several different perspectives.
First of all, the $s+1$ energy levels of the DQHO are equally spaced
as in the QHO. Secondly, the raising and lowering operators $D^{\dagger}$
and $D$ for the DQHO approach to their counterparts in the QHO as
the dimension $s\rightarrow\infty.$ For a fixed $n,$ when $s\rightarrow\infty,$
equations (\ref{eq:eng_raise}) and (\ref{eq:eng_lower}) become
\begin{eqnarray}
\lim_{s\rightarrow\infty}D^{\dagger}|n,s\rangle & = & 2\sqrt{s}\left(1+O\left(\frac{n}{s}\right)\right)\sqrt{n+1}|n+1,s\rangle,\label{eq:eng_raise-1}\\
\lim_{s\rightarrow\infty}D|n,s\rangle & = & 2\sqrt{s}\left(1+O\left(\frac{n}{s}\right)\right)\sqrt{n}|n-1,s\rangle.\label{eq:eng_lower-1}
\end{eqnarray}
Up to an insignificant normalization constant, these are the energy
ladder equations of the QHO. 

As another justification for identification of the DQHO, we show how
the energy eigenstate wave functions of the QHO are recovered by the
DQHO. In the QHO, the energy eigenstate wave functions are, by definition,
the projections of the energy eigenstates on the position eigenstates.
Following this definition, the wave function of the $n$-th energy
eigenstate in the DQHO is the projection of $|n,s\rangle$ on $\psi_{n_{1},s}$,
\begin{equation}
\alpha_{n_{1},s}^{n}\equiv\langle\psi_{n_{1},s}|n,s\rangle.\label{eq:wave-function}
\end{equation}
To calculate $\alpha_{n_{1},s}^{n}$, we re-express Eq.\,(\ref{eq:wave-function})
as

\begin{equation}
|n,s\rangle=\sum_{n_{1}=0}^{s}\alpha_{n_{1},s}^{n}\psi_{n_{1},s}.
\end{equation}
Algorithmically, eigenstates $|n,s\rangle$, specified by $\alpha_{n_{1},s}^{n}$,
may be constructed by applying the $B_{1}^{\dagger}$ operators, which
raises the resolution $s$ by one without affecting the energy $n$,
to the $\psi_{0,0}\equiv|0,0\rangle$ state. Performing 
\begin{equation}
(B_{1}^{\dagger})^{s}\psi_{0,0}=\sqrt{(2s)!!}|n=0,s\rangle,\label{eq:highest_eng}
\end{equation}
gives the lowest energy eigenstate in the $d$-dimensional basis (representation),
while applying the $B_{2}^{\dagger}$ operator to the $\psi_{0,0}$
state $s$ times gives the highest energy state in the $d$-dimensional
basis. The other energy-resolution eigenstates $|n,s\rangle$ in this
$d$-dimensional basis may be found by applying the $D$ and $D^{\dagger}$
operators. This procedure provides an algorithmic definition of $\alpha_{n_{1},s}^{n}$. 

To see how $\alpha_{n_{1},s}^{n}$ recovers the energy eigenstate
wave functions of the QHO, we note that $\alpha_{n_{1},s}^{n}$, can
be expressed as a known special function called Wigner's little-d
function, which can be rewritten in terms of the weighted Kravchuk
function as well as the ordinary hypergeometric function,
\begin{align}
\alpha_{n_{1},s}^{n} & =\text{d}_{-n+s/2,\,n_{1}-s}^{s/2}\left(\frac{\pi}{2}\right)=K_{n}(n_{1},s)\nonumber \\
 & =(-1)^{n}2^{-s/2}\binom{s}{j}^{1/2}\binom{s}{n_{1}}^{1/2}{}_{2}F_{1}(-n,-n_{1};-s;2).\label{eq:alpha}
\end{align}
Note that this is not the standard notation for the Kravchuk functions
in terms of the Kravchuk polynomials, as discussed in Appendix B.
The weighted Kravchuk functions form a complete orthonormal set on
the $n_{1}$ space for fixed $s$, are known to limit to the Hermite
functions for a fixed $n$ as $s\rightarrow\infty$, and obey recurrence
relations which limit to those of the Hermite functions for a fixed
$n$ as $s\rightarrow\infty$ \citep{nikiforov91}. However, for any
fixed $s$, it is only possible to construct $\lfloor s/2\rfloor$
states with continuous analogs. This may be readily understood from
the fact that each increase in energy of the Hermite functions adds
a zero-crossing to the wavefunction. When the number of zero-crossings
exceeds $\lfloor s/2\rfloor$, the wavefunctions will experience aliasing.
Thus, in order to form a complete set, the Kravchuk functions for
$n\ge s/2$ cannot limit to the Hermite functions even as $s\rightarrow\infty$.

That the Kravchuk functions and Wigner's little-d function become
the Hermite functions for fixed $n$ as $s\rightarrow\infty$ is well
known \citep{lorente01}. It is because of this relationship that,
up to a rescaling of variables, these wavefunctions are the starting
point for previous Kravchuk DQHO work \citep{atakishiev90,lorente97}.
But from the algebraic DQHO formalism developed in the present study,
the discrete wave functions of the energy eigenstates are rigorously
defined and constructed without resorting to these special functions,
while still recovering them. The raising and lowering operators for
energy, position, and resolution serve as a sufficient and effective
algorithm for calculating the wave functions of the DQHO. 

This bring us to the following interesting observation. Using the
algebraic DQHO formalism developed, we are able to rigorously derive
the energy eigenstate wave functions of the QHO in a purely algebraic
manner without using differential equations or differential operators.
This is impossible in the continuous or infinite discrete setting
and attests to the usefulness of the algebraic DQHO. 

\begin{figure}
\includegraphics[scale=0.5]{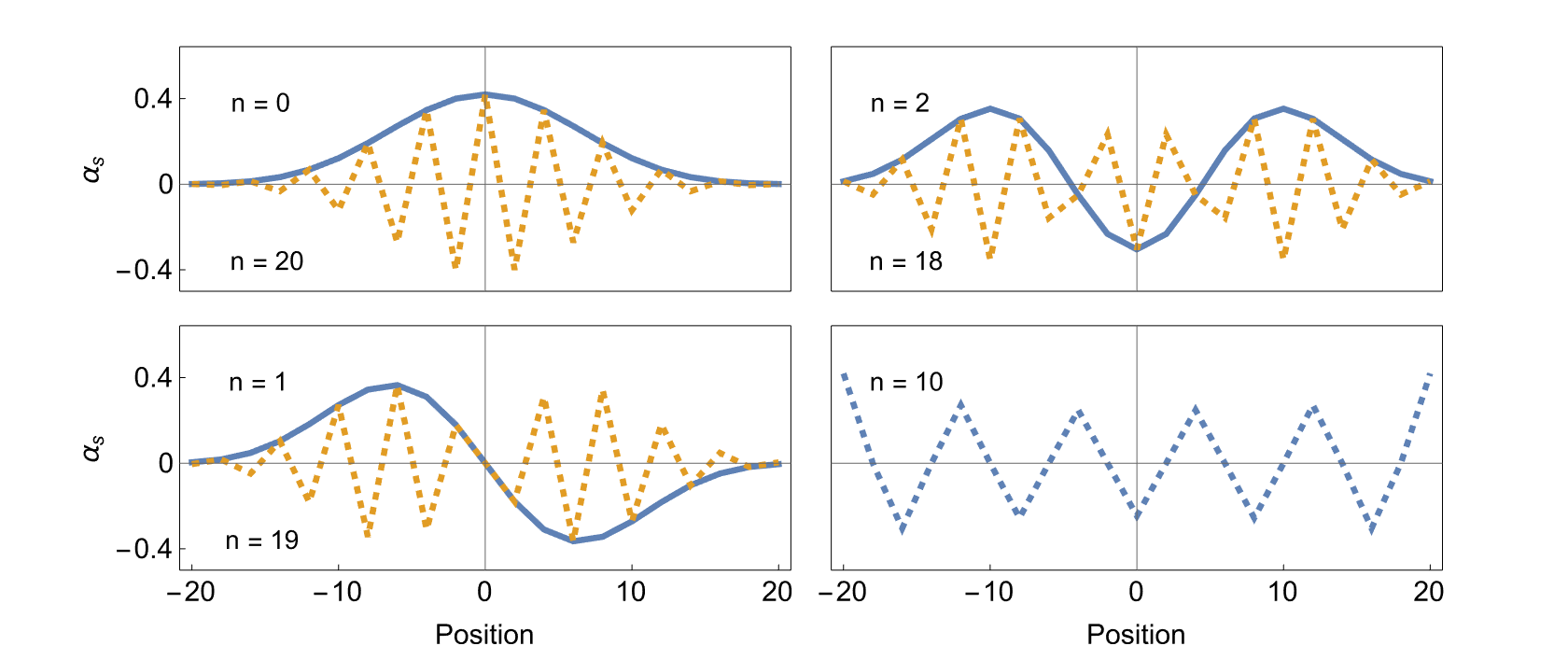}\caption{Representative wave functions of the algebraic DQHO given by energy-resolution
eigenstate $\alpha_{n_{1},s}^{n}$ for $s=20$ and $n=0,1,2,10,18,19,20$.
The vertical axes are $\alpha_{n_{1},s}^{n}$ and the horizontal axes
are position $x=n_{1}-n_{2}$. Although each wavefunction consists
of a set of discrete points, they are drawn as lines for clarity.
The lowest energy eigenstate $\alpha_{n_{1},20}^{0}$ (solid blue
curve) corresponds to the ground state of the QHO. Modes with $n<s/2$
(solid blue curves) furnish finite degree-of-freedom approximations
of the bounded QHO eigenmode wave functions with resolution of $d=s+1=21$.
Energy modes with $n\ge s/2$ (dashed orange curves) cannot be reached
from the ground state in the limit $d\rightarrow\infty$. They form
another kind of DQHO called inverse DQHO that has no continuous counterpart.}

\label{fig:1}
\end{figure}

A set of wave functions $\alpha_{n_{1},s}^{n}$ are shown in Fig.\,\ref{fig:1}
for $s=20$ and $n=0,1,2,10,18,19,20$. Although the $n=j$ and $n=s-j$
states are orthogonal (as all the energy eigenstates are), they have
the same amplitude at each position. But the relative phase between
adjacent positions are different for these two states. The higher
energy state is more oscillatory then the lower energy one. Note that
the algebraic DQHO allows an $n=s/2$ mode when $d=s+1$ is an odd
number. Unlike the $n<s/2$ and $n>s/2$ energy modes, which have
the same expectation values as the QHO wave functions in the limit
$d\rightarrow\infty$, the $n=s/2$ mode has no QHO analog. Indeed,
this energy eigenstate will always extend from $\psi_{0,s}$ to $\psi_{s,s}$,
corresponding to an unrenormalizable wave function in the continuous
limit. Since increasing the energy of the highest $n<s/2$ energy
state, $|s/2-1,s\rangle$, will require the DQHO to pass through this
unrenormalizable state, we can interpret the spectrum as consisting
of two different oscillators. The lower energy states with $n<s/2$
furnish a DQHO that corresponds to the QHO. The higher energy states
with $n>s/2$ form another DQHO that does not have a known continuous
counterpart. For easy reference, we will refer to it as the inverse
DQHO. 

The unrenormalizable energy states have not been identified in previous
Kravchuk DQHO literature. Indeed, the significance of the $n=s/2$
state may be lost, especially if the energy eigenvalues are shifted
by a constant value as in other literature where, although the eigenstates
are the same, the energy eigenvalues may shifted by a factor of $1/2$
to look more similar to those of the QHO \citep{atakishiev90,lorente97}.
To the authors' knowledge, the existence of the inverse DQHO and the
unrenomalizable $n=s/2$ state has not been addressed before.

\section{Coherent States\label{sec:Coherent-States}}

For the QHO, coherent states are 1) eigenstates of the energy lowering
operator, 2) obtained via application of the displacement operator
to the ground state, and 3) minimum uncertainty states. However, in
a finite basis, the first and third of these definitions become problematic.
The first definition requires that the QHO have an infinite number
of energy eigenstates, otherwise the energy lowering operator would
eliminate the highest energy eigenstate. Thus, in a finite setting,
the eigenstates of the energy lowering operator are trivial. The third
definition also poses a problem in this finite setting. We find that
since either the variance in position $\sigma_{X}^{2}$ or the variance
in momentum $\sigma_{P}^{2}$ can be zero,
\begin{equation}
\min\sigma_{X}^{2}\sigma_{P}^{2}=0.
\end{equation}
The minimum uncertainty states are just the union of the position
and momentum eigenstates. 

We will thus proceed with the second definition of coherence. Several
different displacement operators have been proposed with the intention
of preserving the displaced-Gaussian shape of the QHO in the finite,
discrete setting \citep{wolf07,uri23}; however, since the definition
of the QHO displacement operator 
\begin{equation}
\hat{\mathcal{D}}(\beta)=\text{exp}\left(\beta D^{\dagger}-\beta^{*}D\right)
\end{equation}
can be easily translated to our finite system, there is no \emph{ab
initio }reason to modify it. Here, $\beta$ is a displacement parameter.
Therefore, in the present study, the coherent state of the DQHO is
constructed as
\begin{equation}
|\beta\rangle=\hat{\mathcal{D}}(\beta)|0,s\rangle.\label{eq:beta}
\end{equation}

There is significant literature on the application of the displacement
operator to elements of the $\mathfrak{su}(2)$ algebra which result
in \emph{spin coherent states} \citep{radcliffe71,arecchi72}. However,
this literature is solely concerned with the energy-resolution basis,
so discussions of the $n_{1}$, $s$ basis have been limited to work
related to DQHOs. For the Kravchuk DQHO, expectation values of the
Hamiltonian and energy ladder operators have been found for displacement
operator coherent states \citep{draganescu09}, but discussion and
calculation of the wavefunctions and position expectation values of
these states has not been presented before. 

We have now rigorously proven that the expected position of the DQHO
coherent state $|\beta\rangle$ constructed in Eq.\,(\ref{eq:beta})
oscillates as a classical harmonic oscillator, and that $|\beta\rangle$
recovers the coherent state of the QHO when the resolution $s\rightarrow\infty.$ 

\begin{figure}
\includegraphics[scale=0.5]{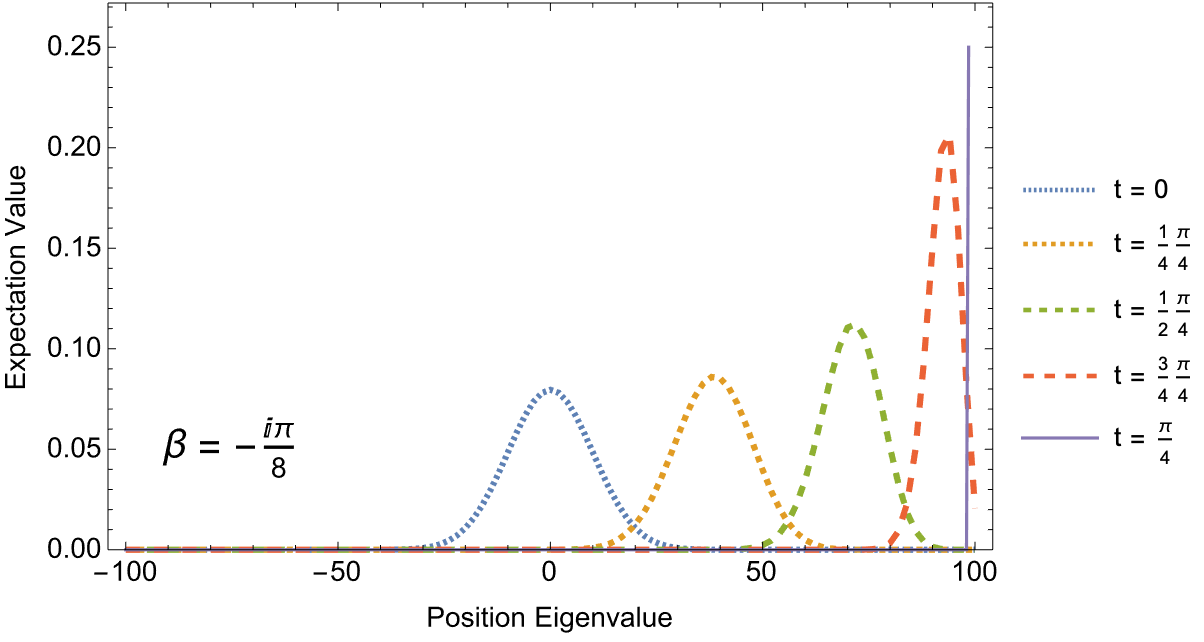}\caption{Expectation value of the position eigenstates of the DQHO coherent
state $\hat{\mathcal{D}}(\beta)|0,s=100\rangle$ . All plots are for
$\beta=i\pi/2$, and $t$ ranges from $0$ to $\pi/4$. The initial
state, $t=0$, corresponds to the ground state $n=0$ of Fig. \ref{fig:1}
(albeit at a different resolution).}

\label{fig:coherent_fixed_beta}
\end{figure}

Using the operator matrix notation and calculation given in Appendix
\ref{sec:Explicit-Matrix-Representations}, we find that in the $n_{1}$,
$s$ basis, 
\begin{align}
\text{log\ensuremath{\left(\hat{\mathcal{D}}(\beta)\right)}}=\  & \text{band}_{s+1\times s+1}\left[1\times1\rightarrow\left(\beta-\beta^{*}\right)\left(s-2j\right)_{j=0}^{s},\right.\nonumber \\
 & \qquad\qquad\qquad\left.1\times2\rightarrow-\left(\beta+\beta^{*}\right)\left(\sqrt{s-j}\sqrt{j+1}\right)_{j=0}^{s-1}\right.,\nonumber \\
 & \qquad\qquad\qquad\left.2\times1\rightarrow\left(\beta+\beta^{*}\right)\left(-\sqrt{s-j}\sqrt{j+1}\right)_{j=0}^{s-1}\right].
\end{align}
In the case that the displacement parameter is imaginary, $\beta=i|\beta|$,
we find a simple, diagonal expression for the displacement operator:
\begin{equation}
\hat{\mathcal{D}}(i|\beta|)=\text{band}_{s+1\times s+1}\left[1\times1\rightarrow\left(e^{2i|\beta|(s-2j)}\right)_{j=0}^{s}\right].
\end{equation}
Note that with $\beta=i|\beta|$, there is no initial displacement
of state, and the displacement only occurs with time evolution. It
is straightforward to calculate the expectation value of the position
operator for a coherent state using the energy-resolution basis. For
a particular resolution, $X$ is a sum of energy raising and lowering
operators, and $U=e^{-iHt}$ is diagonal. Using the Baker-Campbell--Hausdorff
(BCH) identity, the displacement operator can be transformed to a
finite sum of energy raising operators when acting on the ground state
since,
\begin{equation}
\hat{\mathcal{D}}(\beta)|0,s\rangle\propto e^{\frac{\beta}{2|\beta|}\text{tan}(2|\beta|)D^{\dagger}}|0,s\rangle.\label{eq:bch}
\end{equation}
Note that this differs from the typical application of the BCH identity
for the QHO because the DQHO energy raising and lowering operators
obey a different commutation relation, $[D,D^{\dagger}]=8H_{I}$,
from those of the QHO, $[a,a^{\dagger}]=1$ \citep{truax85}. We find
after much simplification (see Appendix \ref{sec:Appendix-D:-Coherent}),
\begin{equation}
\langle\beta|X|\beta\rangle=s\ \text{cos}(2t-\text{arg}(\beta))\ \text{sin}(4|\beta|).\label{eq:coherent_x_expectation}
\end{equation}
Note that the time dependence comes from moving from the Heisenberg
picture operator $X$ to the Schrödinger picture operator $X_{S}=e^{-iHt}Xe^{iHt}$.
The maximum amplitude is simply $s$, as expected, and the expectation
value oscillates harmonically. A quarter period of the wavefunction
in the $n_{1}$, $s$ basis for such a maximum amplitude oscillation
may be seen in Fig. \ref{fig:coherent_fixed_beta}. 

\begin{figure}
\includegraphics[scale=0.5]{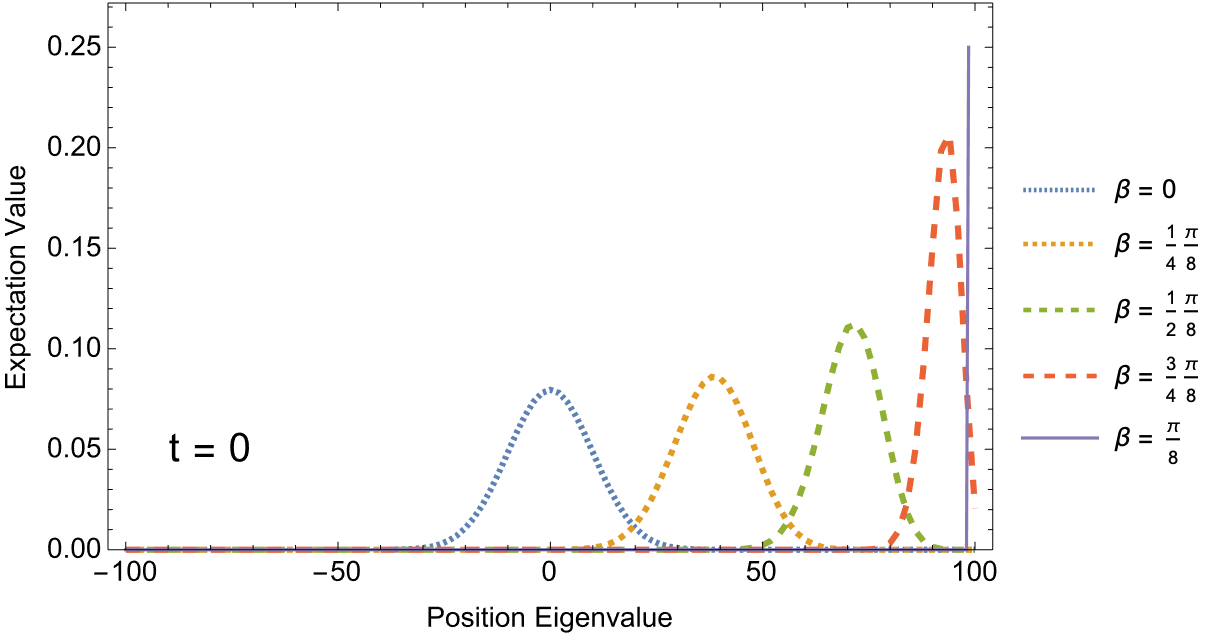}\caption{Expectation value of the position eigenstates for different values
of initial displacement $\beta$ of the DQHO coherent state $\hat{\mathcal{D}}(\beta)|0,s=100\rangle$.
All plots are for $t=0$, and $\beta$ ranges from $0$ to $\pi/8$.
The ground state, $\beta=0$, corresponds to the ground state $n=0$
of Fig. \ref{fig:1} (albeit at a different resolution). The position
eigenstate amplitudes are identical to those in Fig. \ref{fig:coherent_fixed_beta}.}

\label{fig:coherent_fixed_time}
\end{figure}

As suggested by Eq. (\ref{eq:coherent_x_expectation}), there is a
symmetry between the argument of $\beta$ and $t$, and the expectation
value of the wavefunction is unaffected by the transformation $t^{\prime}=-\text{arg}(\beta)/2$,
$\text{arg}(\beta^{\prime})=-2t$. Indeed, it happens that the amplitude
of the wave function is only a function of the wavefunction's average
position as demonstrated in Figure \ref{fig:coherent_fixed_time}.
Thus, fixing time $t=0$ and $\text{arg}(\beta)=0$, we find that
varying the initial maximum displacement $|\beta|$ results in the
same wavefunctions as we found by varying time for a fixed displacement.

Correspondence between the continuous QHO and the DQHO can be found
in the limit that $|\beta|\ll1$. Here, the amplitude of the oscillation
is linearly proportional to $|\beta|$, and the wavefunction, as seen
in Figs. \ref{fig:coherent_fixed_beta} and \ref{fig:coherent_fixed_time}
is minimally deformed from its initial state. 

\section{Discussion\label{sec:Discussion}}

In addition to providing a new and simple derivation of the $\mathfrak{su}(2)$
oscillator based on two coupled quantum harmonic oscillators, which
may be of intrinsic interest, the algebraic DQHO formalism developed
in the present study also creates a formal structure for connecting
representations of DQHOs at different dimensions via the resolution
raising and lowering operators. The Kravchuk oscillator has never
been simulated on a quantum computer, as noted in the introduction,
and the resolution operators provide additional impetus for this to
change. Especially now, while quantum compute time is scarce, the
ability to dynamically alter the resolution of a simulation could
be decisive. Increasing the resolution of a simulation would also
improve its fault tolerance in finer gradations than those available
in typical quantum error correction schemes. For example, suppose
we have some number of qubits available for computation, $Q$, and
we want to simulate a DQHO of dimension $d$. We require $Q\ge\text{log}_{2}(d)$
to minimally perform the simulation, so we have the difference of
qubits left for fault tolerance. Most error correcting schemes require
additional qubits at some fixed proportion $r$ to the dimension of
the simulation, so $Q\ge\text{log}_{2}(d)+\text{log}_{2}(r)$. However,
for the DQHO described here, we may increase the dimension of the
simulation in steps of any size $p\ge1$, naturally entangling the
information over more qubits. Thus, for the DQHO, $Q\ge\text{log}_{2}(d+p)$.
Of course, this is not a substitute for error correction, but since
it may be deployed at any scale, it can almost always act as a supplement. 

Since every computational operation on a quantum computer must be
unitary, nonunitary operators, such as the resolution raising and
lowering operators, need to be embedded in higher dimensional unitary
operators to act on a quantum computer. This embedding is always possible,
but it may require the dimension of the problem to be (temporarily)
doubled. Future work as it relates to quantum computation could involve
a search for efficient unitary embeddings of each of the DQHO's nonunitary
operators. 

Finally, although the DQHO has never been simulated on a quantum computer,
the minimally nonlinear quantum three-wave interaction, governed by
$H_{TW}=igA_{1}^{\dagger}A_{2}A_{3}-ig^{*}A_{1}A_{2}^{\dagger}A_{3}^{\dagger}$
has been simulated \citep{shiPhysRevA,Shi2017,Shi2018}. Note that
the constant $g$ is a coupling coefficient which may be taken to
$-i$ without affecting the dynamics. Harmonic behavior has been previously
found in the limit of $A_{2}\rightarrow\infty$ or $A_{3}\rightarrow\infty$
\citep{may_qin}. In the latter case, the quantum three wave interaction
Hamiltonian just becomes the interaction term $H_{I}$ defined by
Eq.\,(\ref{eq:Hi}) multiplied by the constant $\sqrt{n_{2}}$. In
the three-wave interaction, this represents amplitude exchange of
two small waves mediated by a third, much larger wave. 

\begin{acknowledgments}
This research was supported by the U.S. Department of Energy (DE-AC02-09CH11466). 
\end{acknowledgments}

\section*{Appendix A: Normalizations of energy and resolution operators}

To calculate the normalization of the energy and resolution operators,
we use a method a method similar to that for the QHO. We will begin
with the energy lowering operator $D=B_{1}^{\dagger}B_{2}$. From
Eqs.\,(\ref{eq:dqho_ham}), (\ref{eq:S}), and (\ref{eq:bc}), see
that 
\[
H=\frac{1}{2}B_{2}^{\dagger}B_{2}
\]
and
\[
S=\frac{1}{2}\left(B_{1}^{\dagger}B_{1}+B_{2}^{\dagger}B_{2}\right).
\]
Subtracting these, we have $B_{1}^{\dagger}B_{1}=2(S-H)$, and since
$[B_{1},B_{1}^{\dagger}]=2$,
\begin{equation}
B_{1}B_{1}^{\dagger}=2(S-H+1).\label{eq:b_bdag}
\end{equation}
Next, we calculate the squared normalization of $B_{1}^{\dagger}$:
\[
\left|\mathcal{N}_{B_{1}^{\dagger}}\right|^{2}=\langle n,s|B_{1}B_{1}^{\dagger}|n,s\rangle=\langle n,s|2(S-H+1)|n,s\rangle=2(s-n+1).
\]
Similarly, the squared normalization of of $B_{2}$ may be found:
\[
\left|\mathcal{N}_{B_{2}}\right|^{2}=\langle n,s|B_{2}^{\dagger}B_{2}|n,s\rangle=2n.
\]
Choosing the arbitrary phase factors to be 1, we find the normalization
factor of \textbf{$B_{1}^{\dagger}$} to be 
\[
\mathcal{N}_{B_{1}^{\dagger}}=\sqrt{2(s-n+1)}
\]
and of $B_{2}$ to be
\[
\mathcal{N}_{B_{2}}=\sqrt{2n}.
\]
Next, recall that $B_{1}^{\dagger}$ acts as an $s$ raising operator
while keeping $n$ constant. Also, note that $[S+H,B_{2}]=-2B_{2}$
and $[S-H,B_{2}]=0$, so $B_{2}$ acts as an $s+n$ lowering operator
while keeping $s-n$ constant. So, putting these operators together
we can see that since $D=B_{1}^{\dagger}B_{2}$ will first decrease
$s+n$ by $2$ (while keeping $s-n$ constant) and then increase $s$
by $1$, it acts as an $n$ lowering operator while keeping $s$ constant.
Its normalization will just be the normalization of its constituent
parts, 
\[
D|n,s\rangle=B_{1}^{\dagger}B_{2}|n,s\rangle=2\sqrt{s-n+1}\sqrt{n}|n-1,s\rangle,
\]
as written in Eq. (\ref{eq:eng_lower}). The calculations for the
normalizations of $D^{\dagger}$ and $B_{1}$ proceed in the same
way. Since $B_{1}$ acts as an $s$ lowering operator,
\[
\mathcal{N}_{B_{1}}=\sqrt{2(s-n)},
\]
and since $B_{2}^{\dagger}$ acts as an $s+n$ raising operator,
\[
\mathcal{N}_{B_{2}^{\dagger}}=\sqrt{2(n+1)}.
\]
The operators $B_{1}$ and $B_{2}^{\dagger}$ may then be applied
to a representative state function $|n,s\rangle$ to find the normalization
of $D^{\dagger}$.

\section*{Appendix B: Kravchuk Functions}

The normalized Kravchuk functions are generally written in terms of
\[
k_{n}(x;p,N)=(-1)^{n}{N \choose n}p^{n}{}_{2}F_{1}(-n,-x;-N;p^{-1}),
\]
the Kravchuk polynomials, where $0<p<1$, $N\in\mathbb{N}$, and $n\in\{0,1,\dots,N\}$.
The prefactors of the hypergeometric functions are usually omitted
in older literature (\citep{Koornwinder82}). The Kravchuk function
is then defined in terms of the polynomials as 
\[
K_{n}^{(p)}(x,N)=d_{n}^{-1}\sqrt{\rho(x)}k_{n}(x;p,N)
\]
with the normalization 
\[
d_{n}^{2}={N \choose n}(p(1-p))^{n}
\]
and the weighting (in order to ensure orthogonality) 
\[
\rho(x)={N \choose x}p^{x}(1-p)^{N-x}.
\]
To move from this common notation to that used in Eq. (\ref{eq:alpha}),
we take $p=1/2$ and suppress it as an argument, $N=s+1$, and $x=n_{1}$. 

\section*{Appendix C: Matrix Representations of the Oscillator Operators\label{sec:Explicit-Matrix-Representations}}

In the position-resolution basis, each operator is naturally written
with three indices, the first describing its dimension and the second
two describing the operator's effect at that dimension. For simplicity,
we will consider the operators acting on a $d=s+1$ fixed-dimensional
state so that we can suppress the first index and write each operator
as a matrix. With these assumptions, the $A_{1}$, $A_{2}$, $A_{1}^{\dagger}$,
and $A_{2}^{\dagger}$ operators each have a single non-zero band.
$A_{1}$ is an $s\times s+1$ matrix with nonzero elements along its
$1\times1$ diagonal:
\begin{equation}
A_{1}=\left(\begin{array}{c}
\\
\\
\\
\\
\end{array}\right.\underbrace{\begin{array}{ccccc}
\sqrt{s} & 0 & 0 & \dots & 0\\
0 & \sqrt{s-1} & 0 & \dots & 0\\
\vdots & \vdots & \ddots &  & 0\\
0 & 0 &  & 1 & 0
\end{array}}_{s+1}\left.\begin{array}{c}
\\
\\
\\
\\
\end{array}\right)\left.\begin{array}{c}
\\
\\
\\
\\
\end{array}\right\} s\ .
\end{equation}
Let us define a compact notation for such matrices $M=\text{band}_{p\times q}\left[l\times m\rightarrow(f_{j})_{j=0}^{N}\right]$,
where $p\times q$ indicates the size of the matrix, $l\times m$
indicates the location of the first entry of the nonzero band, and
$(f_{j})_{j=0}^{N}$ is the sequence which populates that band. In
this notation, 
\begin{equation}
A_{1}=\text{band}_{s\times s+1}\left[1\times1\rightarrow\left(\sqrt{s-j}\right)_{j=0}^{s-1}\right].
\end{equation}
Writing the other operators in this notation, we have
\begin{align}
A_{2}= & \text{band}_{s\times s+1}\left[1\times2\rightarrow\left(\sqrt{j+1}\right)_{j=0}^{s-1}\right],\\
A_{1}^{\dagger}= & \text{band}_{s+2\times s+1}\left[1\times1\rightarrow\left(\sqrt{s+1-j}\right)_{j=0}^{s}\right],\\
A_{2}^{\dagger}= & \text{band}_{s+2\times s+1}\left[2\times1\rightarrow\left(\sqrt{j+1}\right)_{j=0}^{s}\right].
\end{align}
Note that in this notation, $A_{j}^{\dagger}$ is not just the Hermitian
conjugate of $A_{j}$ since $A_{j}^{\dagger}$ increases a state's
dimension, while $A_{j}$ decreases it. The same will be true of $F$
and $F^{\dagger}$. 

We can generalize this notation to band diagonal matrices with multiple
nonzero bands. Thus, 
\begin{equation}
H_{I}=-\frac{1}{2}\text{band}_{s+1\times s+1}\left[1\times2\rightarrow\left(\sqrt{s-j}\sqrt{j+1}\right)_{j=0}^{s-1},2\times1\rightarrow\left(\sqrt{s-j}\sqrt{j+1}\right)_{j=0}^{s-1}\right].
\end{equation}
The energy lowering and raising operators become 
\begin{align}
D=\  & \text{band}_{s+1\times s+1}\left[1\times1\rightarrow\left(s-2j\right)_{j=0}^{s},1\times2\rightarrow\left(\sqrt{s-j}\sqrt{j+1}\right)_{j=0}^{s-1},\right.\nonumber \\
 & \qquad\qquad\qquad\left.2\times1\rightarrow\left(-\sqrt{s-j}\sqrt{j+1}\right)_{j=0}^{s-1}\right]\\
D^{\dagger}=\  & \text{band}_{s+1\times s+1}\left[1\times1\rightarrow\left(s-2j\right)_{j=0}^{s},1\times2\rightarrow\left(-\sqrt{s-j}\sqrt{j+1}\right)_{j=0}^{s-1},\right.\nonumber \\
 & \qquad\qquad\qquad\left.2\times1\rightarrow\left(\sqrt{s-j}\sqrt{j+1}\right)_{j=0}^{s-1}\right],
\end{align}
The position operator is simply the diagonal matrix
\begin{equation}
X=\text{band}_{s+1\times s+1}\left[1\times1\rightarrow\left(s-2j\right)_{j=0}^{s}\right],
\end{equation}
while the momentum operator is
\begin{align*}
P= & \text{band}_{s+1\times s+1}\left[1\times2\rightarrow\left(-i\sqrt{s-j}\sqrt{j+1}\right)_{j=0}^{s-1},\right.\\
 & \qquad\qquad\qquad\left.2\times1\rightarrow\left(i\sqrt{s-j}\sqrt{j+1}\right)_{j=0}^{s-1}\right].
\end{align*}

\section*{Appendix D: Coherent State Position Expectation Value\label{sec:Appendix-D:-Coherent}}

We seek to calculate $\langle\beta|X(t)|\beta\rangle$, which can
be expanded to 
\begin{equation}
\langle\beta|X(t)|\beta\rangle=\left\langle 0,s\right|\hat{\mathcal{D}}(\beta)e^{iHt}X_{s}e^{-iHt}\hat{\mathcal{D}}(\beta)\left|0,s\right\rangle ,
\end{equation}
where our $X(t)$ in the Heisenberg picture is transformed to $e^{iHt}X_{s}e^{-iHt}$
in the Schrödinger picture. We will begin by noting that 
\begin{equation}
\hat{\mathcal{D}}(\beta)\left|0,s\right\rangle =\frac{1}{\mathcal{N}^{1/2}}e^{\frac{\beta}{2|\beta|}\text{tan}(2|\beta|)D^{\dagger}}|0,s\rangle,
\end{equation}
as noted in Eq. (\ref{eq:bch}), where $\mathcal{N}$ is some normalization.
Because we have a finite basis, this exponential can be expanded as
a finite sum,
\begin{align}
\hat{\mathcal{D}}(\beta)\left|0,s\right\rangle = & \frac{1}{\mathcal{N}^{1/2}}\Bigg(|0,s\rangle\\
 & \left.+\sum_{j=1}^{s}\left(\frac{\beta}{|\beta|}\text{tan}(2|\beta|)\right)^{j}\frac{1}{j!}\prod_{k=0}^{j-1}\left[2\sqrt{s-k}\sqrt{1+k}\right]|k+1,s\rangle\right),
\end{align}
where we've made use of the normalizations of Eqs. (\ref{eq:eng_lower})
and (\ref{eq:eng_raise}). The product may rewritten using factorial
notation, 
\begin{equation}
\prod_{k=0}^{j-1}\left[2\sqrt{s-k}\sqrt{1+k}\right]=2^{j}\sqrt{j!}\sqrt{\frac{s!}{(s-j)!}},
\end{equation}
so that we have 
\begin{align}
\hat{\mathcal{D}}(\beta)\left|0,s\right\rangle = & \frac{1}{\mathcal{N}^{1/2}}\Bigg(|0,s\rangle\\
 & \left.+\sum_{j=1}^{s}\left(\frac{2\beta}{|\beta|}\text{tan}(2|\beta|)\right)^{j}\sqrt{\binom{s}{j}}|j,s\rangle\right).\label{eq:displaced_expanded_simplified}
\end{align}
Finding the normalization $\mathcal{N}$ simply requires calculating
\begin{equation}
\left|\hat{\mathcal{D}}(\beta)\left|0,s\right\rangle \right|^{2}=1,
\end{equation}
so
\begin{equation}
\mathcal{N}=1+\sum_{j=1}^{s}\text{tan}^{2j}(2|\beta|)4^{j}\binom{s}{j},
\end{equation}
which simplifies to
\begin{equation}
\mathcal{N}=\left(1+4\text{tan}^{2}(2|\beta|)\right)^{s}.
\end{equation}

Because we are working the in the energy-resolution basis, multiplying
Eq. (\ref{eq:displaced_expanded_simplified}) by $U(t)=e^{-iHt}$
is simple. Each state $|n,s\rangle$ is simply multiplied by $e^{-int}$:
\begin{align}
e^{-iHt}\hat{\mathcal{D}}(\beta)\left|0,s\right\rangle = & \frac{1}{\mathcal{N}^{1/2}}\Bigg(e^{ist}|0,s\rangle\\
 & \left.+\sum_{j=1}^{s}e^{i(s-2j)t}\left(\frac{2\beta}{|\beta|}\text{tan}(2|\beta|)\right)^{j}\sqrt{\binom{s}{j}}|j,s\rangle\right).\label{eq:u_d_psi}
\end{align}
Next, multiplying by position operator, $X_{s}=\frac{1}{2}\left(D^{\dagger}+D\right),$
we have
\begin{align}
X_{s}e^{-iHt}\hat{\mathcal{D}}(\beta)\left|-s,s\right\rangle = & \frac{1}{\mathcal{N}^{1/2}}\Bigg(\sqrt{s}e^{ist}|-s+2,s\rangle\\
 & +\sum_{j=1}^{s}e^{i(s-2j)t}\left(\frac{2\beta}{|\beta|}\text{tan}(2|\beta|)\right)^{j}\sqrt{\binom{s}{j}}\label{eq:x_u_d_psi-1}\\
 & \times\left[\sqrt{s-j}\sqrt{j+1}|j+1,s\rangle\right.\\
 & \left.+\sqrt{s-j+1}j^{1/2}|j-1,s\rangle\right]\Bigg).
\end{align}
Further multiplying by $e^{iHt}$ and simplifying, 
\begin{align*}
e^{iHt}X_{s}e^{-iHt}\hat{\mathcal{D}}(\beta)\left|0,s\right\rangle = & \frac{1}{\mathcal{N}^{1/2}}\Bigg(\frac{2\beta}{|\beta|}\text{tan}(2|\beta|)se^{-2it}|0,s\rangle\\
 & +\sum_{j=1}^{s}\left(\frac{2\beta}{|\beta|}\text{tan}(2|\beta|)\right)^{j-1}\sqrt{\binom{s}{j}}|j,s\rangle\\
 & \times\left[je^{2it}+4\left(\frac{\beta}{|\beta|}\text{tan}(2|\beta|)\right)^{2}\left(s-j\right)e^{-2it}\right]\Bigg).
\end{align*}
Finally, we can calculate
\begin{align*}
\langle\beta|X(t)|\beta\rangle= & \frac{1}{\mathcal{N}}\Bigg(\frac{2\beta}{|\beta|}\text{tan}(2|\beta|)se^{-2it}+\sum_{j=1}^{s}2^{2j-1}\frac{\beta^{*}}{|\beta|}\text{tan}^{2j-1}(2|\beta|)\binom{s}{j}\\
 & \times\left[je^{2it}+4\left(\frac{\beta}{|\beta|}\text{tan}(2|\beta|)\right)^{2}\left(s-j\right)e^{-2it}\right]\Bigg)\\
= & \frac{1}{\mathcal{N}}2s|\beta|^{-1}\text{tan}(2|\beta|)\left(\beta e^{-2it}+\beta^{*}e^{2it}\right)\left(1+4\text{tan}^{2}(2|\beta|)\right)^{s-1}\\
= & 2s|\beta|^{-1}\text{tan}(2|\beta|)\left(\beta e^{-2it}+\beta^{*}e^{2it}\right)\left(1+4\text{tan}^{2}(2|\beta|)\right)^{-1}\\
= & 4s\ \text{tan}(2|\beta|)\text{cos}(2t-\text{arg}(\beta))\left(1+4\text{tan}^{2}(2|\beta|)\right)^{-1}\\
= & s\ \text{cos}(2t-\text{arg}(\beta))\ \text{sin}(4|\beta|),
\end{align*}
which is the same as Eq. (\ref{eq:coherent_x_expectation}). 

\bibliographystyle{apsrev4-2}
%

\end{document}